\documentclass[aps,prl,twocolumn,showpacs,amsmath,amssymb]{revtex4} 
\usepackage{graphicx} 
\usepackage{dcolumn} 
\usepackage{bm} 

\begin{document}

\title{Multiple transient memories in experiments on sheared non-Brownian suspensions}

\author{Joseph D. Paulsen$^{1,2,4}$}
\email{paulsenj@umass.edu}
\author{Nathan C. Keim$^{3,4}$}
\author{Sidney R. Nagel$^{4}$}
\affiliation{\mbox{$^1$Department of Physics, University of Massachusetts, Amherst, Massachusetts 01003, USA}
\mbox{$^2$Department of Polymer Science and Engineering, University of Massachusetts, Amherst, Massachusetts 01003, USA}
\mbox{$^3$Department of Mechanical Engineering and Applied Mechanics, University of Pennsylvania, Philadelphia, Pennsylvania 19104, USA}
\mbox{$^4$James Franck Institute and Department of Physics, The University of Chicago, Chicago, Illinois 60637, USA}}

\date{\today}

\begin{abstract}
A system with multiple transient memories can remember a set of inputs but subsequently forgets almost all of them, even as they are continually applied. 
If noise is added, the system can store all memories indefinitely. 
The phenomenon has recently been predicted for cyclically sheared non-Brownian suspensions. 
Here we present experiments on such suspensions, finding behavior consistent with multiple transient memories and showing how memories can be stabilized by noise. 
\end{abstract}


\maketitle

A physical system has memory if it is endowed with the basic operations of imprinting, retrieval, and erasure. 
Common examples are mechanical marking or the flipping of magnetic domains. 
More exotic examples include return-point memory~\cite{Barker1983,Sethna1993} and aging and rejuvenation in glasses~\cite{Jonason1998,Zou2010}. 
These systems all support the intuition that (i) the more times an input is presented the stronger the memory becomes, and (ii) random noise is detrimental to memory retention. 
However, both attributes are violated by multiple transient memories, which have been seen in traveling charge-density waves~\cite{Coppersmith1997,Povinelli1999} and predicted for sheared non-Brownian suspensions~\cite{Keim2011,Keim2013PRE}. 
The experiments reported here on sheared suspensions demonstrate that noise can stabilize this form of memory retention. 

Keim and Nagel~\cite{Keim2011} described how multiple transient memories could occur in a simplified model of a suspension under cyclic shear: 
When sheared repeatedly between strain amplitudes $\gamma=0$ and $\gamma=\gamma_1$, a suspension can organize into a reversible steady state, thereby encoding a memory of $\gamma_1$. 
The memory appears as a sudden drop in reversibility as the strain amplitude is swept past $\gamma_1$. 
Multiple memories can be formed if several amplitudes, $\gamma_1<\gamma_2<...<\gamma_n$, are repeatedly applied. 
However, once the suspension relaxes to a state that is completely reversible up to amplitude $\gamma_n$, it is also reversible for all $\gamma<\gamma_n$; thus the memories of all the smaller training amplitudes are effectively erased. 
The presence of noise was predicted to prevent the system from reaching a fully reversible state so that other memories could be retained. 

For multiple transient memories in charge-density waves, the role of noise was only demonstrated in a simulation~\cite{Povinelli1999}; in experiments~\cite{Coppersmith1997} the ambient noise could not be varied and was assumed to be strong enough so that the system could remember all inputs. 
In the present paper, we cyclically shear neutrally buoyant, non-Brownian suspensions at low Reynolds number. 
By varying the noise, we demonstrate explicitly that noise is required to retain a memory of all input strain amplitudes at long times. 
This provides a concrete example of the emergence of plasticity in memory.

\textit{Experiment.}---In the experiment, a viscous suspension is cyclically sheared in a $6.3$ mm gap between two cylinders in a circular Couette geometry (with an inner cylinder radius of $36.6$ mm). 
The suspension is composed of PMMA spheres (Cospheric, LLC) in a mixture of Triton X-100, water, and zinc chloride (dynamic viscosity $\mu=$ 4,300 mPa s) that is index and density matched to the PMMA particles following ref.~\cite{Krishnan1996}. 
Except where otherwise stated, the particles have diameters, $d$, between $d=106$ and $125$ $\mu$m, with volume fraction of $\phi=0.35$. 
The suspension is slowly sheared by rotating the inner cylinder, keeping the Reynolds number (the ratio of inertial to viscous forces in the fluid) below $Re=0.007$. 
The P\'eclet number (the ratio of advection to diffusion) is $\sim10^9$ so that the particles are effectively non-Brownian. 
The suspension floats on a low-viscosity ($\mu=24$ mPa s) fluorinated oil (Fluorinert FC-70, 3M Company) and is open to air above so that the bottom and top surfaces are essentially stress-free. 
Before each experiment, the particle locations are randomized by applying one or two $360^{\circ}$ rotations.

\begin{figure}[tb]
\centering 
\begin{center} 
\includegraphics[width=3.4in]{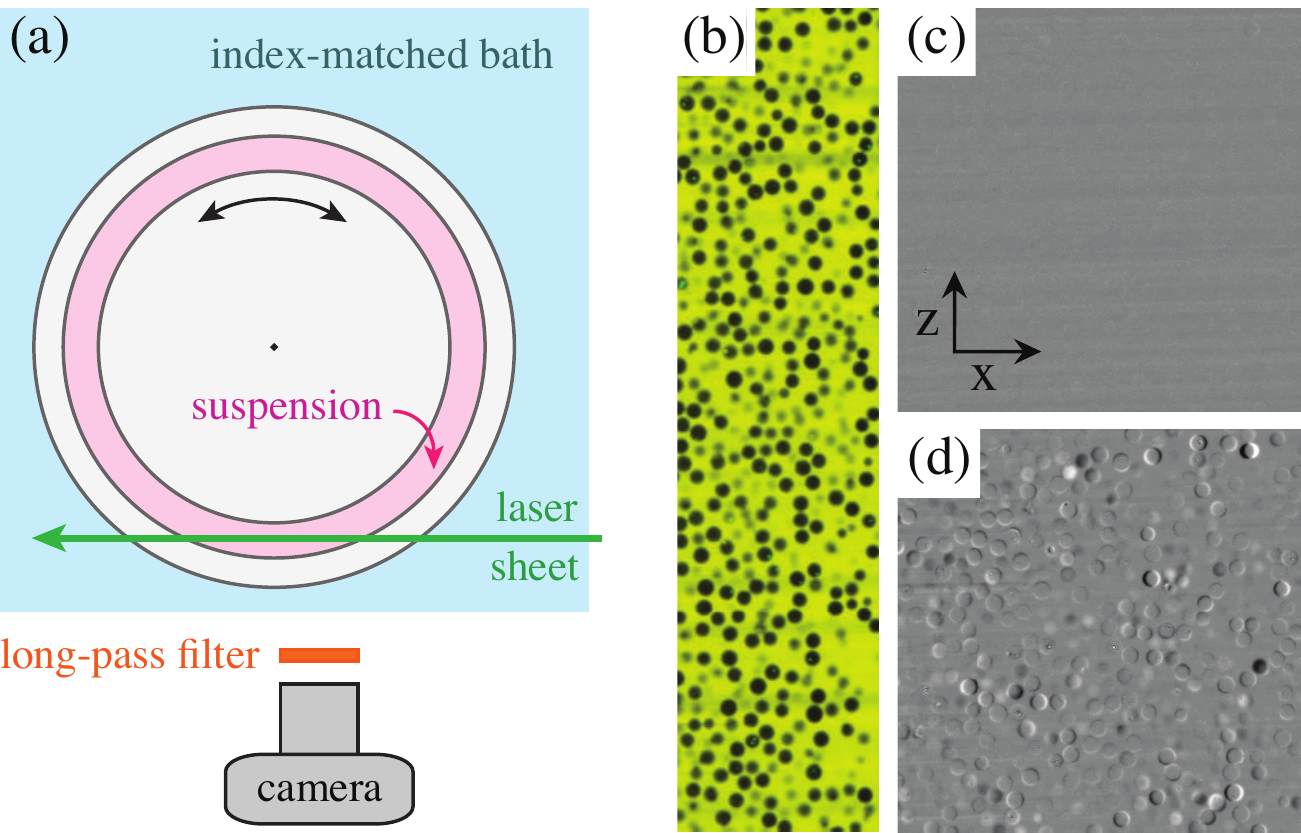} 
\end{center}
\caption{
(color online). Experimental setup. 
(a) Top view of the circular Couette cell containing a viscous suspension between two concentric cylinders. 
A two-dimensional slice of the suspension is imaged by shining a laser sheet into the fluorescently-dyed fluid. 
The emitted light is imaged through a long-pass filter.
(b) A small region of the imaged slice.
(c, d) Visual readout of a memory formed at $\gamma_1=1.2$. 
Each image is the difference of pictures taken before and after a single back-and-forth rotation of amplitude (c) $\gamma=1.2$ and (d) $\gamma=1.4$. 
The subtractions show that the particle trajectories are reversible at $\gamma=1.2$ but irreversible at $\gamma=1.4$. 
}
\label{Apparatus}
\end{figure}

Fluorescent dye (Rhodamine 6G) is added to the fluid so that a two-dimensional slice of the suspension can be imaged using a laser sheet ($\lambda=532$ nm). 
By submerging the cell in an index-matched bath, as shown in Fig.~\ref{Apparatus}(a), the laser sheet is not refracted as it enters the cell. 
Following~\cite{Pine2005,Corte2008}, we image the suspension stroboscopically, taking one picture at $\gamma = 0$ for each cycle. 
Part of the field of view is pictured in Fig.~\ref{Apparatus}(b). 
By comparing successive images, we identify the degree of reversibility of the suspension. 
If the particle trajectories are completely reversible, then the two images will be identical. 

\textit{Single memories.}---Previous experiments~\cite{Pine2005,Corte2008} had shown that, starting from a random configuration, the particle trajectories are initially irreversible but eventually reach a configuration where they retrace their paths exactly during each cycle. To demonstrate single-memory formation, we shear an initially randomized suspension cyclically between $\gamma=0$ and $\gamma_1=1.2$ for 200 cycles. 
A readout consists of applying a series of back-and-forth rotations of increasing strain amplitude, from $\gamma=0$ to $\gamma=3$ in increments of $\Delta\gamma=0.2$. 
Figure~\ref{Apparatus}(c,d) shows how this protocol detects a memory. In Fig.~\ref{Apparatus}(c), the image taken immediately before shearing by amplitude $\gamma=1.2$ is subtracted from the one taken immediately after. 
The result is approximately monotone, indicating that the particle trajectories are nearly reversible. 
Figure~\ref{Apparatus}(d) shows the subtraction for the next shear, $\gamma=1.4$. 
The particles are now clearly displaced, revealing a memory of amplitude $1.2\leq \gamma <1.4$.

\begin{figure}[tb]
\centering 
\begin{center} 
\includegraphics[width=3.4in]{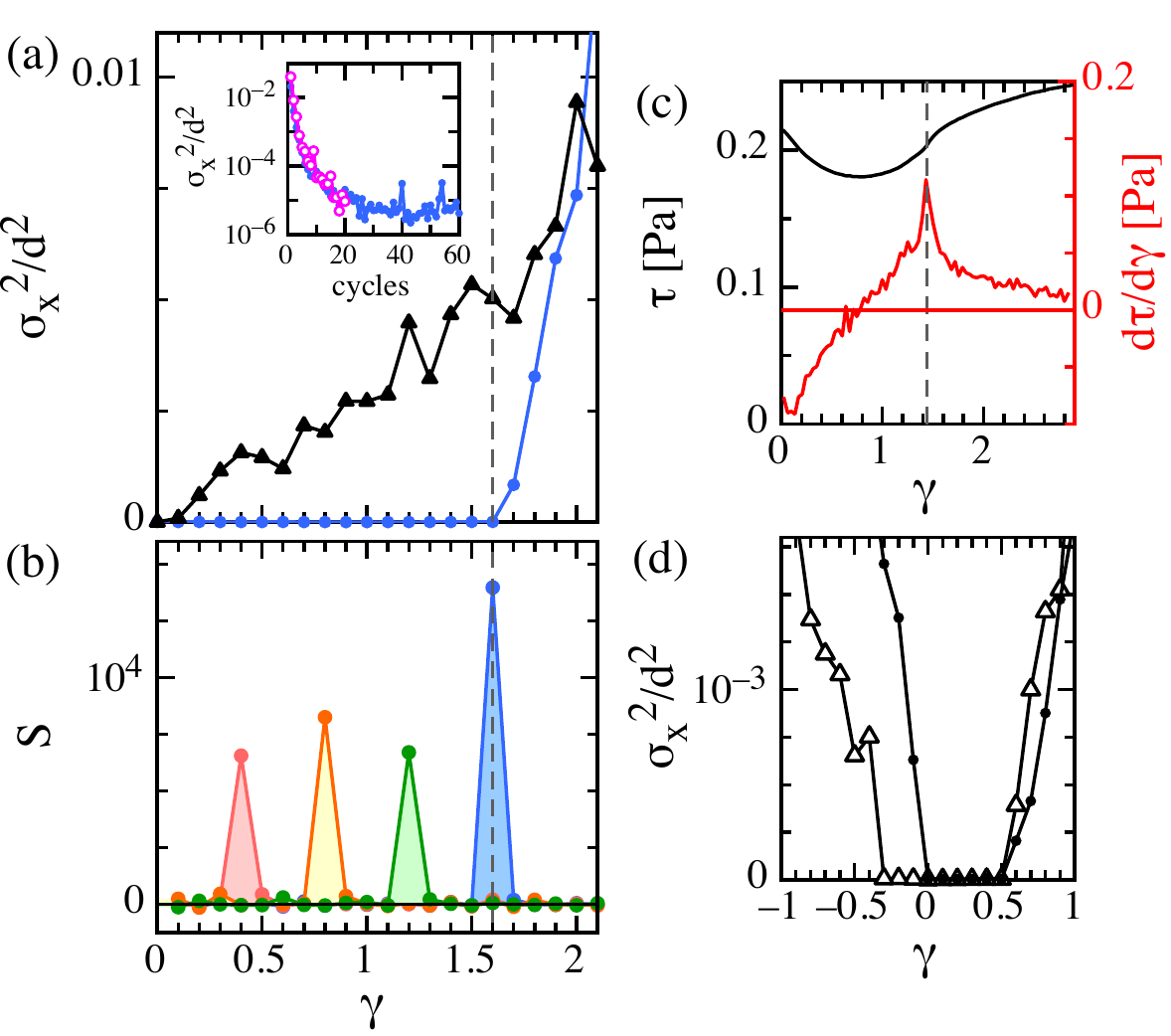} 
\end{center}
\caption{
(color online). 
Single memories. 
(a) \textit{Inset}: $\sigma_x^2/d^2$ versus cycle number for $\gamma_1=1.6$. 
The system relaxes to a reversible steady-state in $\sim$30 cycles. 
Closed circles: $Re=0.007$. 
Open circles: 20 cycles with $Re=0.001$. 
\textit{Main}: Memory readout. 
$\sigma_x^2/d^2$ versus readout strain, $\gamma$. 
Circles: Readout after training with 100 cycles of $\gamma_1=1.6$. 
The suspension is reversible up to $\gamma_1$. 
Triangles: Readout for a randomized suspension shows no memory. 
(b) $S$ (defined by eqn.~\ref{Sequation}) versus readout strain, for systems trained for 100 cycles at $\gamma_1=0.4$, $0.8$, $1.2$, and $1.6$.
The peaks identify the memory values. 
(c) Rheology of a single memory showing the stress versus strain during a readout shear. 
After training with 10 cycles of $\gamma_1= 1.44$, the stress ($\tau$, left axis) on the inner cylinder is measured during a unidirectional constant strain-rate shear ($\dot{\gamma}=0.018$ s$^{-1}$).
The stress sharply increases at $\gamma_1=1.44$ (dashed line), where there is a peak in the slope of the data ($d\tau/d\gamma$, right axis), indicating the memory.
Here $Re=0.0002$, $d=90$ to $106$ $\mu$m, and inner cylinder radius = $13.3$ mm.  
(d) Two sides to a single memory: $\sigma_x^2/d^2$ versus readout strain for single memories, showing readouts in both the $+$ (clockwise) and $-$ (anticlockwise) directions (with $\phi=0.45$). 
Suspensions were trained between $\gamma=0$ and $\gamma_{1}=0.5$ (circles), and between $\gamma_{1-}=-0.3$ and $\gamma_{1+}=0.5$ (triangles). 
}
\label{SingleMem}
\end{figure}

In order to isolate {\em relative} particle displacements as opposed to uniform drifts, we track particles~\cite{Crocker1996} to measure the variance of their displacements in the $x$ direction after a cycle, normalized by the square of the particle diameter: $\sigma_x^2/d^2$. If the particle paths are completely reversible, $\sigma_x^2/d^2=0$. 
The inset to Fig.~\ref{SingleMem}(a) shows $\sigma_x^2/d^2$ versus cycle number for an initially randomized system that is sheared repeatedly to $\gamma_1=1.6$. 
To check that the experiments are in the low Reynolds-number limit, we repeated the experiments at two shear rates corresponding to $Re=0.007$ and $Re=0.001$. 
The inset to Fig.~\ref{SingleMem}(a) shows that the behavior is the same at the two speeds. 

We now examine the readout of a single memory, which has been trained by applying 100 cycles of $\gamma_1=1.6$. 
Figure~\ref{SingleMem}(a) shows $\sigma_x^2/d^2$ versus readout amplitude. 
To increase resolution, we interleave the data from two experiments (each with $\Delta\gamma=0.2$, but one starting at $\gamma=0$ and the other starting at $\gamma=0.1$). 
There is a sharp increase in $\sigma_x^2/d^2$ at $\gamma_1=1.6$, thus identifying the memory formed there. 
(The memory is present in the $z$-component of the variance as well, although the readout is more noisy.) 
 
To highlight the memory, we define a signal, $S$, as: 
\begin{equation}
S\equiv (\sigma_x^2)''/\sigma_x^2, 
\label{Sequation}
\end{equation}
where prime ($'$) denotes a derivative with respect to $\gamma$. 
A sharp peak in $S$ indicates a memory. 
To show that the memory value can be freely chosen, in Fig.~\ref{SingleMem}(b) we plot $S$ for systems that were trained over a range of strains: $\gamma_1=0.4$, $0.8$, $1.2$, and $1.6$. 

A memory can also be seen in rheology. 
Previous work showed that the storage modulus (averaged over a single shear cycle) decays during the approach to a reversible steady state~\cite{Corte2008}. 
Here we show that a memory may be retrieved simply by monitoring the stress, $\tau$, while the suspension is unidirectionally sheared from $\gamma=0$. 
We apply 10 shear cycles at amplitude $\gamma_1= 1.44$, 
and then measure the stress on the inner cylinder during a unidirectional constant strain-rate shear starting at $\gamma=0$. 
As shown in Fig.~\ref{SingleMem}(c), the stress shows a sharp increase at $\gamma= 1.44$, identifying the stored memory. In addition to offering another readout method, this shows that a memory is stored in the interactions of the particles as the strain approaches the training amplitude, $\gamma_1$. 

\textit{Memory and shear direction.}---Reference~\cite{Keim2013PRE} noted that in simulations when a memory is encoded by applying cyclic shear between $\gamma=0$ and $\gamma_1>0$, there will be an increase in particle reversibility when the system is strained in the reverse direction to $\gamma<0$ as well as to $\gamma>\gamma_1$. 
Thus, a single memory stores two values, corresponding to the two reversal points between which the system is cycled. 

We find this symmetry in our experiments. 
Figure~\ref{SingleMem}(d) shows $\sigma_x^2/d^2$ from a system that was trained with $\gamma_1=0.5$. 
The memory was read out (using $\Delta\gamma=0.1$) in the $+\gamma$ direction and in the $-\gamma$ direction in two separate experiments. 
The memory at $\gamma=0$ can be placed in another location; Fig.~\ref{SingleMem}(d) shows the readout curve where we trained the system between $\gamma_{1-}=-0.3$ and $\gamma_{1+}=0.5$.
As before, the memories are present at the two reversal points of the training~\cite{PulseSign}.

\begin{figure}[tb]
\centering 
\begin{center} 
\includegraphics[width=3.2in]{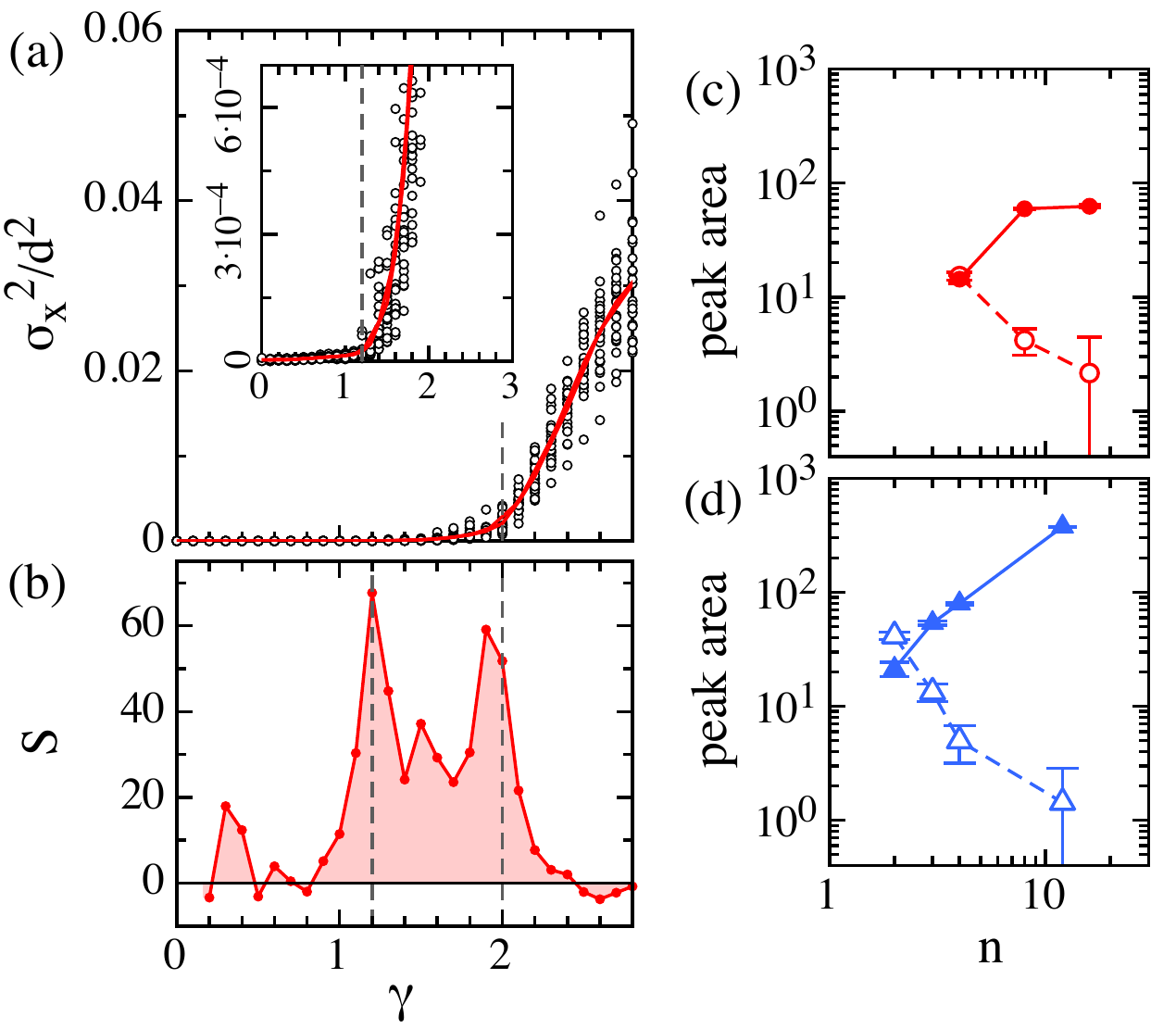} 
\end{center}
\caption{
(color online). 
Multiple Memories. 
(a) Readout of 48 independent experiments with the training sequence $\gamma=2.0,1.2,1.2,1.2,1.2$ repeated 4 times. 
Points: individual runs. 
Line: average values. 
The system shows a sharp increase in irreversibility at the larger training amplitude, $\gamma_2=2.0$. 
\textit{Inset:} A memory of the smaller training value, $\gamma_1=1.2$, is visible on an expanded y-axis.
(b) $S$ of the averaged data in (a).
The two memories are signified by the peaks. 
(c) Area under the peaks in $S$, versus the number of times the training sequence ($\gamma=2.0,1.2,1.2,1.2,1.2$) is applied. 
As the peak at $\gamma_2=2.0$ (closed symbols) becomes stronger, the peak at $\gamma_1=1.2$ (open symbols) gradually disappears until it cannot be resolved from the background (indicated by the error bars), and is effectively forgotten. 
Panel (d) shows similar results for memories at $\gamma_1=0.8$ (open symbols) and $\gamma_2=1.6$ (closed symbols), using the training sequence: $\gamma=1.6,0.8,0.8,0.8,0.8,0.8,0.8$. 
}
\label{MultipleMem}
\end{figure}

\textit{Multiple memories.}---As in the simulations in refs.~\cite{Keim2011,Keim2013PRE}, we have formed multiple memories in our experiments by cyclically applying more than one strain amplitude. 
We repeatedly apply the sequence $\gamma_2,\gamma_1,\gamma_1,\gamma_1,\gamma_1$, where $\gamma_1=1.2$ and $\gamma_2=2.0$.  
Figure~\ref{MultipleMem}(a) shows the readout from 48 independent experiments, where this entire sequence is applied four times. 
The main panel shows a clear increase in particle irreversibility at $\gamma_2=2.0$. 
The inset, where the y-axis of the plot is expanded, shows that the particle irreversibility also increases when the strain exceeds $\gamma_1=1.2$. 
This shows that both memories are stored in the system at the same time.

This is corroborated in Fig.~\ref{MultipleMem}(b), where we plot the signal, $S$, of the averaged data. The two clear peaks correspond to the two memories. 
We expect that if $\gamma_1$ and $\gamma_2$ were applied in equal numbers, the memory of $\gamma_1$ would still be present, although much harder to see. 

As the system continues to be trained, the memory encoded at $\gamma_1$ becomes harder and harder to retrieve while the one at $\gamma_2$ becomes dominant. 
This is because, once the suspension is completely reversible at $\gamma_2$ it is impossible to see any change in reversibility ({\em i.e.}, a memory) at any smaller strain amplitude. 
Thus, while initially it is possible to have a memory of all training amplitudes, the memory of the smaller amplitudes will gradually be erased. 
This effect was predicted by the simulations~\cite{Keim2011,Keim2013PRE}, and we show the corresponding experimental results in Fig.~\ref{MultipleMem}(c,d). 
These figures show the area under the peaks in $S$ at $\gamma_1$ and $\gamma_2$ for the two different training protocols given in the figure caption. 
As $n$, the number of applications of the training sequence, increases, the peak at $\gamma_2$ grows while the one at $\gamma_1$ decreases until it disappears into the background. 

\textit{Noise stabilization of multiple memories.}---In the absence of inertia and any external forces, the suspension should retain a memory indefinitely if undisturbed. 
However, in our experiment, the reversibility gradually erodes as the suspension ages; the memory is robust for a short pause but decays as the pause increases. 
We find that the particle positions drift during the pause, perhaps due to imperfect density matching or small temperature gradients. 
We harness these accumulating perturbations, or `noise,' by introducing a pause after each shear cycle of our training.

\begin{figure}[tb]
\centering 
\begin{center} 
\includegraphics[width=3.0in]{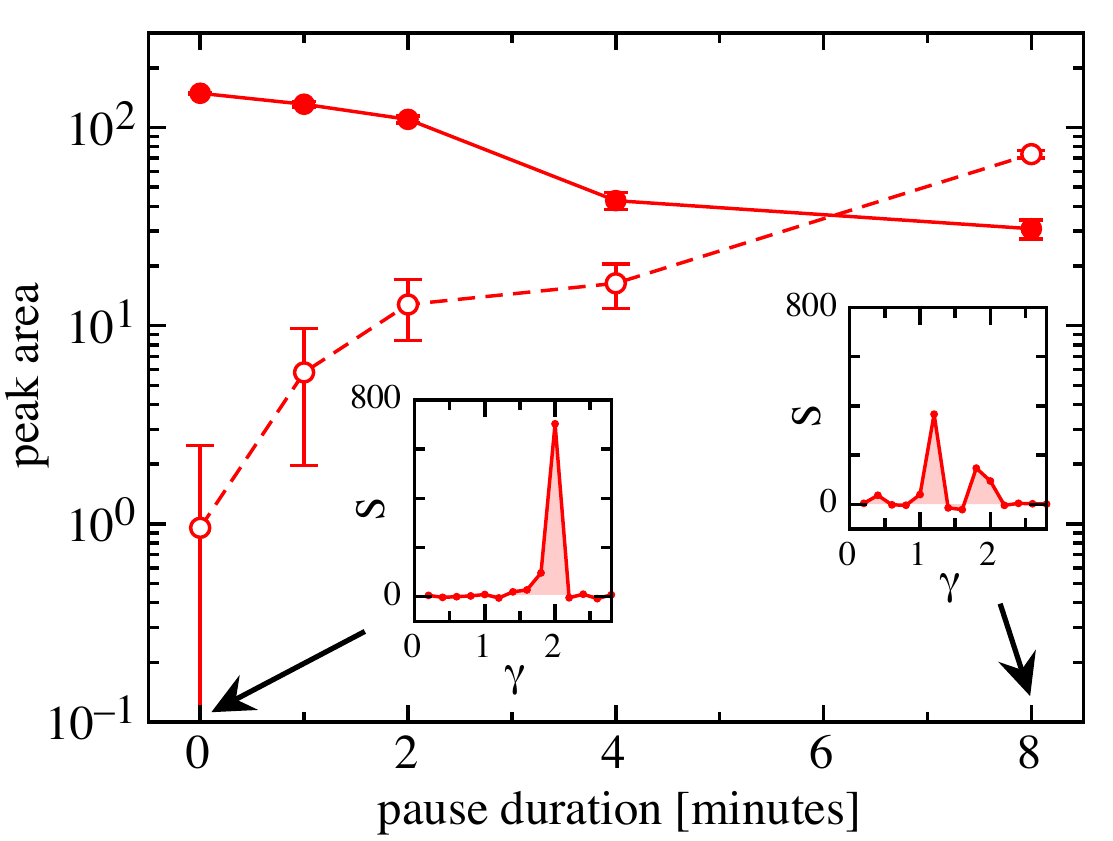} 
\end{center}
\caption{
(color online). 
Stabilization by noise. 
Area under the peaks in $S$ signifying the memories at $\gamma_1=1.2$ (open symbols) and $\gamma_2=2.0$ (closed symbols), versus the duration of the pause between each shear cycle. 
The memory at $\gamma_2=2.0$ is weaker for longer pauses. 
The memory at $\gamma_1=1.2$ cannot be distinguished from the background (indicated by the error bars) for short pauses, but is stabilized by longer pauses. 
The insets show $S$ versus readout strain for 0 and 8 minute pause durations. 
}
\label{ForgetNoise}
\end{figure}

In Fig.~\ref{ForgetNoise}, we show that the presence of this noise can sustain the memory of a smaller input, $\gamma_1$, that would otherwise be overwritten by a large amplitude strain, $\gamma_2$. 
Here, the training sequence $\gamma=2.0,1.2,1.2,1.2,1.2$ is applied 8 times. 
In the inset at the left of Fig.~\ref{ForgetNoise}, there was no pause between cycles, and the memory at $\gamma_1=1.2$ was forgotten. 
In the inset at the right, we apply an identical training protocol, except we now include an 8-minute pause between cycles. 
In this case, $S$ versus $\gamma$ shows {\em both} memories are present: the addition of noise has allowed the smaller memory to survive. 
The main panel of Fig.~\ref{ForgetNoise} shows that as the pause duration between subsequent shear cycles is increased, the peak in $S$ at $\gamma_1$ grows while the peak at $\gamma_2$ shrinks. 
Similar behavior was found in the simulations, where it was interpreted as noise preventing the system from ever reaching a fixed point with complete reversibility up to $\gamma_2$. 

We do not yet know whether the forgetting is sufficiently gradual that one memory always erodes slowly while another takes over. 
In the present experiments with two strain amplitudes, we have not been able to detect the memory at $\gamma_1$ if the larger shear, $\gamma_2$, was the last one applied. 
Gradual forgetting distinguishes multiple transient memories from other classes of memory, such as return-point memory. 
However, simulations of multiple transient memories~\cite{Keim2013PRE} show that if the kick given to the particles during a collision is too large, then the memory of the smaller shear, $\gamma_1$, can be hard to discern, although it is still there and can be detected in large systems or when many averages are taken.  
Indeed, our experiments appear to correspond to this behavior. 
Further experiments should be able to elucidate this issue. 

\textit{Conclusion.}---We have experimentally demonstrated multiple memories in sheared non-Brownian suspensions. 
These have many of the properties of multiple transient memories~\cite{Coppersmith1997,Povinelli1999,Keim2011,Keim2013PRE}:
(i) the suspension can learn multiple memories, (ii) the memory of the smaller input strain is erased even as that input is continually applied, and (iii) the memory of the smaller input value is stabilized by the presence of noise. 
Also, as in charge-density waves, the sheared suspensions remember the direction of the last applied deformation~\cite{PulseSign}. 
It is remarkable that these properties---including the counterintuitive role of noise---emerge in two very different systems.
 
Our results demonstrate an interplay between noise and the transition from irreversible to reversible behavior. 
There must be an optimal amount of noise for effective memory retention: memories are undetectable if noise is too small, and memories can be swamped by noise that is too large. 
However, it is not yet clear how to estimate this optimal noise amplitude or how it depends on the parameters of the system or the values of the inputs to be stored in memory. 
This question might apply to all the ways that the system can become irreversible, such as driving past the maximum strain amplitude for self-organization~\cite{Pine2005,Corte2008,Keim2013PRE}. 
 
A coherent understanding and categorization of memory effects in condensed matter is lacking; there is much room to develop this part of the physics literature. 
As argued in ref.~\cite{Keim2011}, similar behavior to multiple transient memories may occur in other particulate systems, such as granular~\cite{Slotterback2012,Ren2013} and amorphous materials~\cite{Keim2013SM,Regev2013,Keim2014,Fiocco2014}.
Simpler forms of memory, such as the Kaiser~\cite{Kurita1979} and Mullins effects~\cite{Schmoller2013}, are known to occur in other materials under cyclic driving. 
However, the ability of noise to support multiple memories is relatively unexplored.

\begin{acknowledgments} 
We thank Eric Brown and Carlos Orellana for assistance with the rheology and Dustin Kleckner for assistance designing the laser sheet. 
J.D.P. gratefully acknowledges funding from the Grainger Foundation Fellowship. 
This work was supported by NSF Grant DMR-1404841.
Use of the Chicago MRSEC Rheometry Facility is gratefully acknowledged. 
\end{acknowledgments}


\begin{thebibliography}{24}
\bibitem[{Pul()}]{PulseSign}
\bibinfo{note}{After the system is subjected to a large-amplitude strain, we
  find a memory of the shear direction---the particle trajectories are more
  reversible if the shear direction is reversed. This is also the direction in
  which it is easier to shear the system, as has been observed in
  suspensions~\cite{GadalaMaria1980} and granular materials~\cite{Toiya2004}.
  It is also reminiscent of pulse-sign memory in charge-density waves, where
  the voltage response of the system is faster if the second pulse shares the
  sign of the first, versus if they have opposite signs~\cite{Fleming1983}.}
\end{thebibliography}

\end{document}